%% file: main.tex
\pdfoutput=1

\documentclass[prd,aps,amsmath,nofootinbib,amssymb,eqsecnum,showkeys,twocolumn]{revtex4-2}

\usepackage{amsmath, latexsym, amssymb, graphicx, color, multirow, bm, braket, ascmac, bbm, outlines, amsfonts}
\usepackage{hyperref}
\usepackage{mathtools}
\usepackage{subcaption}
\usepackage{lineno}
\usepackage{cancel}
\usepackage{dsfont}
\usepackage[table]{xcolor}
\usepackage{tabularx}
\usepackage[T1]{fontenc}
\usepackage[capitalize]{cleveref}
\usepackage{romannum}
\usepackage{ragged2e}
\usepackage{makecell, physics}
\definecolor{nicered}{rgb}{.7,.1,.1}
\definecolor{nicegreen}{rgb}{.1,.5,.1}
\definecolor{darkblue}{rgb}{0,0,.55}
\definecolor{burntorange}{rgb}{0.8, 0.33, 0.0}
\definecolor{bluee}{rgb}{0,0.1,0.9}
\definecolor{crimson}{rgb}{0.86, 0.08, 0.24}
\hypersetup{colorlinks, citecolor=nicegreen,linkcolor=nicered, urlcolor=darkblue}
\usepackage{multirow}
\usepackage{slashed}
\usepackage{comment}
\usepackage{ulem}
\numberwithin{equation}{section}
\usepackage{verbatim}
\usepackage{simplewick}
\newcommand{\RNum}[1]{\uppercase\expandafter{\romannumeral #1\relax}}

\begin{document}
\captionsetup{justification=RaggedRight}
\pagenumbering{arabic}
	\title{The Multi-Messenger Astroparticle Physics: the First Constraint on Light Millicharged Dark Matter via Time-Delay Analysis of GRB GW170817A
		\\
	}
	\author{Wenxing Zhang$^{1,2,3}$, Junle Pei$^{4}$, Xin Zhang$^{5,6}$, Tianjun Li$^{7,8,9}$\textsuperscript{*} }
	\affiliation{\vspace{2mm} \\
		$^1$Department of Physics, Hebei University, Baoding, 071002, China\\
  		$^2$Hebei Key Laboratory of High-Precision Computation and Application of Quantum Field Theory, Baoding 071002, China\\
		$^3$Hebei Research Center of the Basic Discipline for Computational Physics, Baoding 071002, China \\
        $^4$Institute of Physics, Henan Academy of Sciences, Zhengzhou, 450046, China\\
        $^5$Department of Physics, Liaoning Normal University, Dalian 116029, China\\
        $^6$Center for Theoretical and Experimental High Energy Physics, Liaoning Normal University, Dalian 116029, China\\
        $^7$School of Physics, Henan Normal University, Xinxiang 453007, P. R. China \\
        $^8$CAS Key Laboratory of Theoretical Physics, Institute of Theoretical Physics, Chinese Academy of Sciences, Beijing 100190, China \\
        $^9$School of Physical Sciences, University of Chinese Academy of Sciences, No. 19A Yuquan Road, Beijing 100049, China \\}


    \input{abstract.tex}
	\maketitle
        \footnotetext {*Corresponding author email: tli@itp.ac.cn}
    \input{Introduction.tex}

    \input{Light_velocity.tex}

    
    \input{conclusion.tex}

\input{acknowledgement.tex}

    \input{Appendix.tex}
    
    \appendix

	\vspace{0.5cm}	
	\phantomsection
	\addcontentsline{toc}{section}{References}
	\bibliography{Reference}
	\bibliographystyle{utphys}
\end{document}

%% file: abstract.tex
\begin{abstract}

The multi-messenger astroparticle physics provides a new approach to probe the new physics beyond the Standard Model. We propose to probe the light dark matter which can interact with electromagnetic interaction. To be concrete, we derive the new constraint on the millicharged dark matter from the multi-messenger observations of GW170817. In the neutron star merger event GW170817, the first detection of a gamma-ray burst (GRB) delayed by approximately 1.7 seconds relative to the gravitational wave emission was observed. Utilizing this delay, we constrain the parameter space of the millicharged dark matter within the large-scale structure of the Universe. {For dark matter mass below $10^{-15}$ eV, the parameter $\epsilon$ is constrained to be less than $10^{-14}$, representing the most stringent limits achieved to date.}

\end{abstract}

%% file: Introduction.tex
\section{Introduction}

The advent of multi-messenger astronomy has inaugurated a new era in fundamental physics, enabling unprecedented probes of the Universe through the correlated detection of gravitational waves (GWs), photons, and neutrinos. Each messenger carries complementary information, and their combined analysis provides a powerful tool for testing the Standard Model and exploring new physics. In this work, we introduce a novel astroparticle phenomenology framework—multi-messenger astroparticle physics—that leverages temporal correlations between different cosmic messengers to search for particle dark matter (DM).
To be concrete, we propose to probe the light DM which can interact with electromagnetic interaction, for instance, millicharged dark matter, axion, and dark photon, etc. In this paper, we concentrate on the millicharged dark matter. The millicharged particles, which carry a fraction of electron charge, $Q=\epsilon e$, are proves of the violation of charge-quantization hypothesis and predicted by the superstring theories~\cite{Wen:1985qj, Shiu:2013wxa} and grand unified theories~\cite{Pati:1973uk, Georgi:1974my}. The dark sector provides a natural scenario for weakly breaking the $U(1)$ electromagnetic symmetry through interactions with hidden photon fields~\cite{Feng:2009mn,Cline:2012is, Dobroliubov:1989mr}. 
The interaction between millicharged Dark Matter and the Standard Model is mediated by a kinetically mixed dark photon~\cite{Holdom:1985ag, Pospelov:2008zw, Fabbrichesi:2020wbt, Arza:2025cou}, effectively introducing a feeble $U(1)$ coupling. For wave-like dark matter with mass below $10^{-10}$ eV, terrestrial detectors such as the milliQan experiment~\cite{milliQan:2021lne}, colliders~\cite{Prinz:1998ua,Davidson:2000hf,Ball:2020dnx}, fixed-target experiments~\cite{Golowich:1986tj,ArgoNeuT:2019ckq,Kelly:2018brz,Choi:2020mbk,Kelly:2018brz,SHiP:2015vad} and neutrino experiments~\cite{Gninenko:2006fi,Majorana:2018gib,Magill:2018tbb}, face challenges due to the ultralight particle’s macroscopic de Broglie wavelength~\cite{Hui:2021tkt}, necessitating probes via spectral distortions in astrophysical~\cite{Arza:2025cou} or cosmological transient events~\cite{Davidson:2000hf}.
In this paper, we do not address the production mechanisms of millicharged dark matter, but instead propose a novel detection scheme for light dark matter in the Universe. This approach leverages the observed time delay between the short gamma-ray burst (sGRB) and gravitational wave emission from the neutron star merger event~\cite{Kasliwal:2017ngb, Savchenko:2017ffs, Kocevski:2017liw}, GW170817~\cite{LIGOScientific:2017vwq, LIGOScientific:2017zic, LIGOScientific:2017ync}. By treating the ubiquitous millicharged dark matter in the cosmos as a homogeneous medium — analogous to light propagation in water where photon scattering with the medium slows the effective phase velocity — we analyze how interactions between photons and millicharged dark matter could induce a measurable delay in high-energy photon arrival times.

The sGRB 170817A~\cite{Pozanenko:2017jrn, Goldstein:2017mmi, Insight-HXMTTeam:2017zsl, LIGOScientific:2017zic} serves as an ideal candidate for probing light dark matter in the Universe due to two key characteristics:
\begin{itemize}
    \item \textbf{Extreme Propagation Distance:}  The source is located 40 Mpc away from the Earth, making the DM distribution along the path approximately homogeneous. In addition, the vast distance amplifies cumulative effects from photon–dark matter interactions.
    \item \textbf{Breakthrough in Time-Delay Measurement:} The event provided the first precise measurement of the propagation time difference ($\sim$1.7s) between the sGRB and gravitational wave signals, establishing a novel pathway to constrain dark matter parameters via multimessenger time-delay analysis. 
\end{itemize}

The sGRB originated from relativistic jets driven by accretion disks~\cite{Granot:2017tbr, Nakar:2007yr, Berger:2013jza} around black holes formed in the aftermath of compact binary mergers. Considering the nonzero jet breakout time, the actual delay of gamma-ray photons during propagation is actually less than 1.7 seconds. 
However, due to the difficulty in precisely calculating the jet breakout duration, we adopt the most conservative scenario in this work and focus on translating observational time delay into constraints on the millicharge parameter, assuming the photon propagation delay approximates the full 1.7s. 
Crucially, if the true delay is smaller than 1.7s, the derived upper limit on the millicharge parameter would be further strengthened. With this delay, we constrain the millicharge parameter of dark matter within the large-scale structure of the Universe. 
For dark matter mass below $10^{-15}$ eV, the millicharge parameter $\epsilon$ is constrained to be less than about $10^{-14}$, representing the most stringent limits achieved to date.

%% file: Light_velocity.tex
\section{Light velocity in DM medium and Constraints from the sGRB GW170817A}\label{sec::light_velocity}
Given that time-delay effects are significantly enhanced for dark matter with mass below $\mathcal{O}(1)$ eV, we specifically focus on millicharged dark matter within sub-eV mass range.
The light dark matter must be bosonic due to the large average number of particles in a de Brogile volume, $\lambda_{dB}^3=\left(\frac{2\pi}{mv} \right)^3$,
\begin{equation}
    N_{dB}\sim\left(\frac{34 ~\text{eV}}{m} \right)^4 \left(\frac{250~\text{km/s}}{v} \right)^3.
\end{equation}

\begin{figure}[bp]\centering 
	\begin{center}
	
           \centering
		\includegraphics[width=0.5\textwidth]{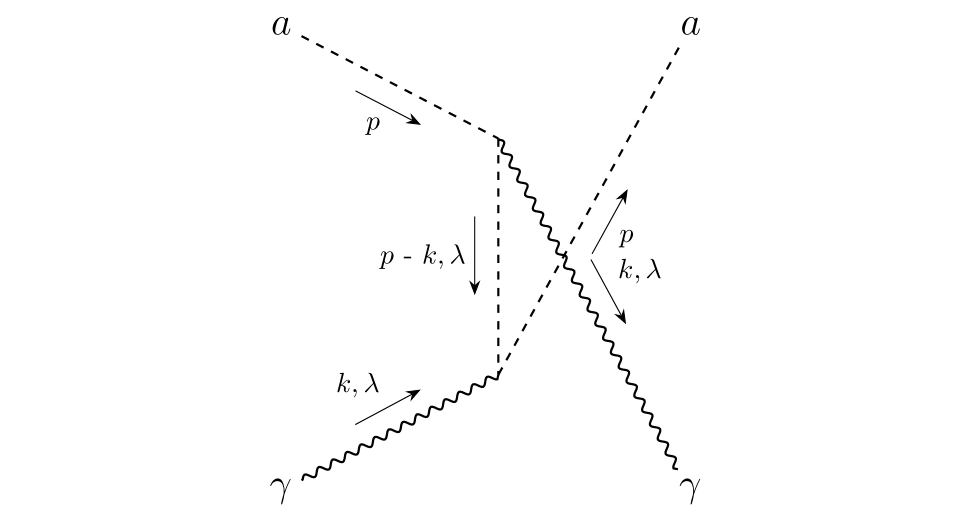}
            \includegraphics[width=0.5\textwidth]{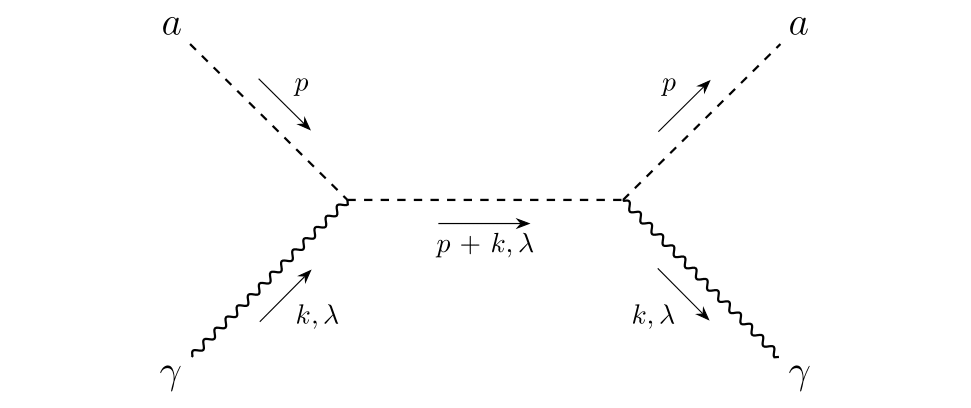}
            \includegraphics[width=0.5\textwidth]{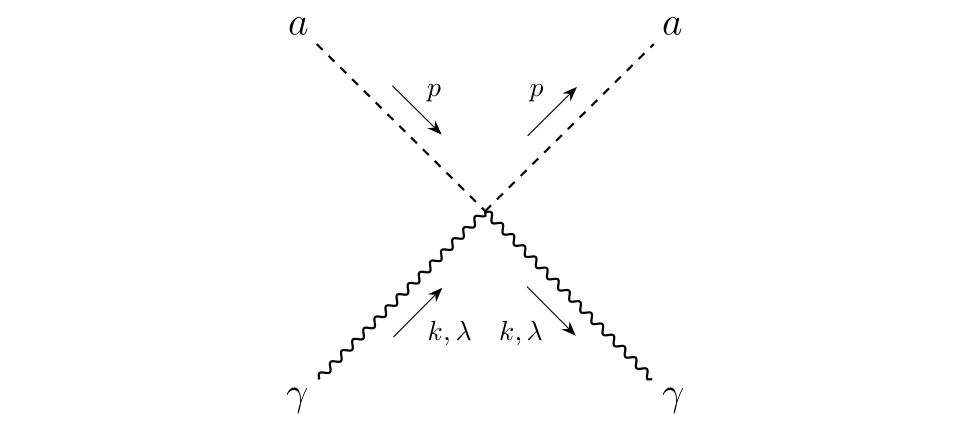}
	\end{center}
	\caption{\label{fig::feyn_diag}Feynman diagrams for the forward scattering process involving gamma photons.}
\end{figure}

The effective refraction index that is relevent with the group velocity for a particle in a homogeneous medium is~\cite{Latimer:2013rja}
\begin{equation}\label{eq::n_index}
    n = 1 + \frac{\rho_N}{4m\omega^2} \mathcal{M} (k, p,\lambda \to k,p,\lambda)
\end{equation}
where $\rho_N=\rho/m$ is the dark matter number density obtained from its relic density ($\rho$) about $0.4~\text{GeV}/\text{cm}^3$, $m$ is dark matter mass, and $\omega$ is photon frequency.
The amplitude, $\mathcal{M} (k, p,\lambda \to k,p,\lambda)$, corresponds to the forward Compton scattering event with $\lambda$ indicating the photon polarization. This value is calculated in the dark-matter rest frame with $p=(m,0)$. One can show that the amplitude vanishes unless the photon polarization direction is unchanged.
For millicharged scalars with electric charge $\epsilon e$, the Lagrangian that is relavent with interactions with SM photon is $\mathcal{L}\supset \left|\mathcal{D}_{\mu}\phi \right|^2$, where $\mathcal{D}_{\mu}=\partial_{\mu}-i\epsilon e A_\mu$ with $\epsilon e$ the scalar millicharge. Three channels including s-, u- and $\phi\phi A_{\mu} A^{\mu}$ direct interaction channels are included as shown in the Fig~\ref{fig::feyn_diag}.
This amplitude is equal to $\left| \mathcal{M} \right| = 2\epsilon^2 e^2$.

One might assume that the DM as the Bose-Einstein condensation~\cite{Sin:1992bg,Ji:1994xh,Hu:2000ke,Evans:2023uxh} would lead to a significant enhancement of scattering amplitudes due to high occupancy within one de-Brogile wave-length volume, rendering single-particle scattering amplitude calculations inapplicable. However, the fact is that this enhancement effect has already been accounted for in refractive index calculations involving multiple scattering. 
We employ the method described in the Ref.~\cite{Evans:2024zlx} to describe dark matter as a multi-particle wave packet, and the result is equivalent to the modification of the propagator presented in the Ref~\cite{Evans:2023uxh}.
To make this explicit, we need to reformulate eq.(\ref{eq::n_index}) as a relation between the time delay and the single-collision amplitude in general
\begin{equation}\label{eq::deltaT}
	\Delta t = \frac{L}{4m \omega^2} \sum\limits_{i} \mathcal{M}_{i} (k, p ,\lambda \to k,p,\lambda),
\end{equation}
where the $\sum_i\mathcal{M}_{i}$ is the amplitude counted for per unit volume. The subscript \textit{i} stands for the \textit{i-th}  forward Compton process within a unit volume. 
$L$ is the distance bewteen the observer and the source.

Similarly, the ultralight dark matter comprises multiple Bose-Einstein condensates (BECs), indexed by i. The \textit{i-th}  BEC consists of a macroscopic number $N_i$ of identical bosonic particles occupying a single quantum state, with their momentum differences lying below experimental resolution.
Such occupation number follows the Bose-Einstein distribution and has the relationship $\rho_N = \sum\limits_{i} N_i$.

A BEC composed of $N_i$ particles—each described by a wave packet—forms a collective background field. When this field participates in scattering processes, the interaction evolves from single-particle scattering to $N_i$-wave-packet scattering, reflecting a macroscopic quantum enhancement.
Our derivation establishes that in Feynman diagrams with two external legs of Axion-Like Particles (ALPs), the BEC-mediated scattering amplitude and the single-particle amplitude obey the relation:
\begin{equation}\label{eq::Mi}
	\mathcal{M}_i = N_i \mathcal{M}(k, p ,\lambda \to k,p,\lambda).
\end{equation}
The detailed calculations are provided in the Appendix. 
From the eq.(\ref{eq::Mi}) and eq.(\ref{eq::deltaT}), one can directly see that the refractive index formula satisfies eq.(\ref{eq::n_index}).
It should be noted that our calculations do not depend on the specific occupation numbers within the coherent state, i.e., they are independent of the explicit form of $N_i$. This is because we address a specialized process that depends solely on the total ALP particle number, and the enhancement factor for individual scattering events coincides exactly with $N_i$.

According to the experimental observations of GW170817A~\cite{Pozanenko:2017jrn, Goldstein:2017mmi}, the sGRB detected by the Fermi satellite exhibited a frequency distribution spanning 8 keV to 300 keV. The time lags between different frequency bins showed no significant frequency dependence~\cite{LIGOScientific:2017vwq, LIGOScientific:2017zic, LIGOScientific:2017ync}. However, a time interval of approximately $0.27 \pm 0.05$s~\cite{Pozanenko:2017jrn} was observed between the arrival times of gamma rays in the (8, 50) keV and (100, 300) keV energy ranges. 
Based on the above results, we consider the following two scenarios:

\begin{itemize}
    \item \textbf{Scenario1:} Neglecting the arrival time differences between different frequency bands and assuming the photon delay approximates the full 1.7s, we adopt the central frequency of 100 keV as the reference frequency. This scenario provides the most conservative constraints on the millicharge parameter. 
    In this case, the millicharge parameter can be expressed as
    \begin{equation}
        \epsilon_{max} \simeq 15.86 \times \left(\frac{m}{1 \text{eV}}\right) \left( \frac{0.4 \text{GeV}/\text{cm}^3}{\rho} \right)^{\frac{1}{2}} \left( \frac{\omega}{100~\text{keV}}\right).
    \end{equation}
    \item \textbf{Scenario2:} Assuming the time lag $\sim 0.27$s between (8, 50) keV and (100, 300) keV photons arriving at Earth is induced by scattering with dark matter. In this case, the sGRB and gravitational waves are not produced simultaneously - the sGRB emission lags behind the GWs by approximately one second, which is consistent with predictions from jet formation models for sGRBs. The time lag induced by the millichared DM is no longer than $\sim 0.27$s. Approximately, adopting $30$ keV and $200$ keV as representative energies for (8, 50) keV and (100, 300) keV bands respectively, we derive
    \begin{equation}
        \epsilon_{max} \simeq 2.26 \times \left(\frac{m}{1 \text{eV}}\right)  \left( \frac{0.4 \text{GeV}/\text{cm}^3}{\rho} \right)^{\frac{1}{2}}.
    \end{equation}

The upper limit of the millicharge parameter derived from the two scenarios is shown in the Fig.~\ref{fig::upper_limit}.
\begin{figure}[bp]\centering 
	\begin{center}
	\begin{subfigure}[t]{0.5\textwidth}
           \centering
		\includegraphics[width=\textwidth]{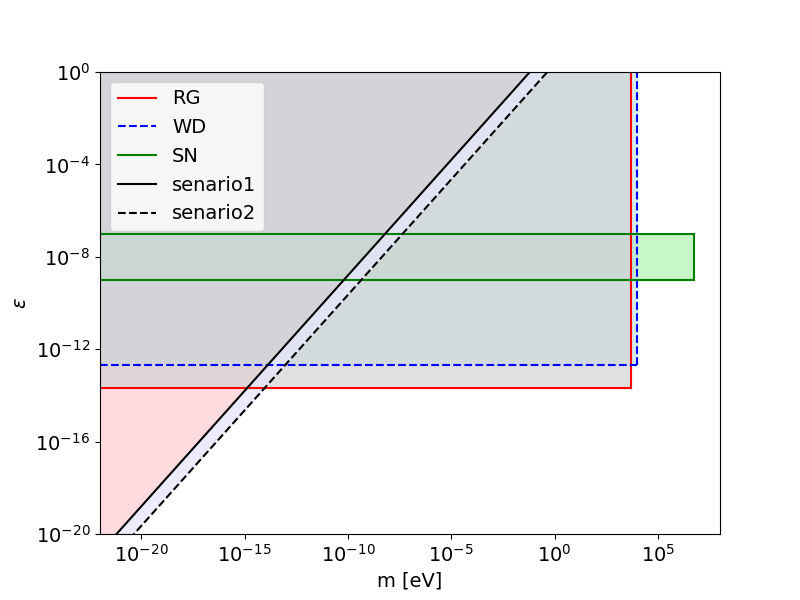}
    \end{subfigure}
	\end{center}
	\caption{\label{fig::upper_limit}Upper limit from cosmological observables from red giant (RG)~\cite{Capozzi:2020cbu}, white drawfs (WD)~\cite{Davidson:1991si}, supernova (SN)~\cite{Chang:2018rso, Fiorillo:2024upk} as well as the sGRB 170817A. The colored area is excluded by the corresponding observables.}
\end{figure}

\end{itemize}


%% file: conclusion.tex
\section{Discussion}\label{sec::discussion}
We have presented a new astroparticle detection strategy that utilizes time-delay measurements between gravitational waves and electromagnetic signals from neutron star mergers to search for light millicharged dark matter. By analyzing the GW170817/GRB 170817A event, we derived competitive constraints on the millicharge parameter $\epsilon$ for ultralight DM—in some mass regimes, surpassing existing limits from laboratory or astrophysical probes.

In this work, we investigate constraints on the millicharged DM deriving from the time delay between the gravitational wave event and its electromagnetic counterpart. To address uncertainties in the relative timing of jet emission and gravitational wave generation during the neutron star merger, we adopt the most conservative scenario by attributing the entire observed 1.7-second delay to dark matter-induced refraction.

By utilizing the connection between the refractive index in a homogeneous dark matter medium and the forward Compton scattering amplitude, our preliminary calculations reveal that GRB 170817A provides the most stringent constraints on $\epsilon$ within the ultralight mass regime ($m_{DM} \lesssim 10^{-15}$eV).

When further incorporating the BEC effects of ultralight dark matter, we demonstrate that—under cosmological large-scale structure conditions—the refractive index modification caused by single-particle Compton scattering amplitudes becomes equivalent to the collective effect of multiparticle BEC condensates. This equivalence enables us to derive the most robust conservative bounds on the millicharged dark matter parameter space.


Our results highlight the potential of multi-messenger timing as a novel experimental tool, complementary to traditional particle detection methods. This approach is inherently cross-disciplinary, bridging gravitational-wave physics, gamma-ray astronomy, and particle phenomenology. Looking ahead, the growing catalog of neutron star mergers from advanced GW detectors will provide a statistically robust dataset to which our method can be applied systematically. Future events with better-resolved emission timelines and broader multi-messenger coverage—including neutrinos—will further sharpen the probe and may even enable discovery.

%% file: acknowledgement.tex
\section{Acknowledgement}
TL is supported in part by the National Key Research and Development Program of China Grant No. 2020YFC2201504, by the Projects No. 11875062, No. 11947302, No. 12047503, and No. 12275333 supported by the National Natural Science Foundation of China, by the Key Research Program of the Chinese Academy of Sciences, Grant NO. XDPB15, by the Scientific Instrument Developing Project of the Chinese Academy of Sciences, Grant No. YJKYYQ20190049, and by the International Partnership Program of Chinese Academy of Sciences for Grand Challenges, Grant No. 112311KYSB20210012.
W. Zhang is supported in part by the National Natural Science Foundation of China under grant no. 12405120, Start-up Funds for Young Talents of Hebei University (No.521100224226), and Funded by Science Research Project of Hebei Education Department QN2026608. J. Pei is supported by the National Natural Science Foundation of China (Project NO. 12505121), by the Joint Fund of Henan Province Science and Technology R$\&$D Program (Project NO. 245200810077), and by the Startup Research Fund of Henan Academy of Sciences (Project NO. 20251820001).

%% file: Appendix.tex
\section{Appendix}

In this section, we will discuss about the Bose-Einstein condension (BEC) effct and demonstrate that the BEC enhancement effect is equivalent to the multiple scattering processes experienced by high-energy photons propagating through a dark matter medium.

We employ multi-particle wave packets to characterize the ingoing or outgoing states of ultralight dark matter particles, where the BEC backgroud field is generated by n wave packets~\cite{Evans:2024zlx}:
\begin{equation}
	|n> = \frac{(a_a^{\dagger})^n}{\sqrt{n!}} |0>
\end{equation}
where the $a_a^{\dagger}$ is the wave packet generating operater that is expressed as
\begin{equation}
	a_{a}^{\dagger} = \int \frac{d^3 p }{(2\pi)^3} \frac{\tilde{a}(p)}{\sqrt{2E_p}} a^{\dagger}(p),
\end{equation}
with the wave function renormalization condition
\begin{equation}
	 \int \frac{d^3 p }{(2\pi)^3} \frac{|\tilde{a}(p)|^2}{2E_p} = 1.
\end{equation}
The field operator $a(x)$ is expressed as
\begin{equation}
	\phi(x)= \int \frac{d^3 p }{(2\pi)^3} \frac{1}{\sqrt{2E_{p_i}}} \left[a(p) e^{-ipx}+a^{\dagger}(p)e^{i p x}\right]
\end{equation}

The amplitude that is relavent with the creation and annihilation of the ALP in the background of BEC fluid can be simplified as 
\begin{align}\label{eq::n_body}
	&~~~~~<N_i| \phi(y)\contraction[1ex]{}{\phi(y)}{}{\phi(x)}\phi(y)\phi(x)\phi(x)~ |N_i>\\\nonumber
	& =\frac{N_i^2}{N_i!}\prod_{j,k=1}^{N_i-1} \int \frac{d^3 p_j}{(2\pi)^3} \int \frac{d^3 q_k}{(2\pi)^3} \frac{\tilde{a}(p_j)\tilde{a}^*(q_k)}{2E_{p_j}2E_{q_k}} <q_k|p_j> \\\nonumber
    & \times	<q| \phi(y)\contraction[1ex]{}{\phi(y)}{}{\phi(x)}\phi(y)\phi(x)\phi(x)~ |p>\\\nonumber
	&= N_i 	<q| \phi(y)\contraction[1ex]{}{\phi(y)}{}{\phi(x)}\phi(y)\phi(x)\phi(x)|p>.
\end{align}

Here, we omit the field operators unrelated to the creation or annihilation of ALPs.
Field operators connected by contraction lines must undergo contraction, while those without contraction symbols participate in the creation and annihilation of initial and final state particles. 
In the second line, the  factor, $N_i^2$, arises from the number of momentum eigenstates p and q selected from the initial and final states.
The last line in the eq.(\ref{eq::n_body}) is obtained by using $<q_k|p_j>=(2\pi)^3 (2E_{p_j})\delta^3(q_k-p_j)$. 
After integrating over the momentum variables $q_k$, this term acquires an $N_i-1!$ factor stemming from the permutation symmetry of the BEC particles' momenta between the initial and final states and finally leaves an enhancement factor of $N_i$.